\documentclass[12pt]{article}
\usepackage{amsmath,epsfig}
\usepackage{amsmath}
\usepackage{amssymb}
\usepackage{mathrsfs}
\usepackage{amsmath,epsfig}
\usepackage{graphicx}
\usepackage{amsmath,epsfig}
\usepackage{epsfig}
\newsavebox{\DSLASH}
\sbox{\DSLASH}{$D$\hspace{-2.5mm}/}
\newcommand{\DS}{\usebox{\DSLASH}}
\begin{document}

\vspace{2cm}
\begin{center}
{\large \bf{Noncommutative Standard Model in Top Quark Sector}}\\
\vspace{2cm}

Mojtaba Mohammadi Najafabadi \footnote{Email:
mojtaba@mail.ipm.ir}\\
{\sl School of Particles and Accelerators, \\
Institute for Research in Fundamental Sciences (IPM)\\
P.O. Box 19395-5531, Tehran, Iran}\\

\vspace{3cm}
 \textbf{Abstract}\\
 \end{center}

In this article we aim to estimate the bounds on the
noncommutative scale $\Lambda_{NC}$ and to extract the $95\%$
exclusion contours for some $\theta_{\mu\nu}$ components using the
recent measurements of the top quark width and the $W$ boson
polarization in top pair events from CDF experiment at Tevatron.

\newpage
\section{Introduction}

The Standard Model (SM) of the particles has been found to be in
good agreement with the present experimental data in many
of its aspects. However, in the framework of the SM top quark is
the only quark which has a mass in the same order as the
electroweak symmetry breaking scale, $v\sim 246$ GeV, whereas all
other observed fermions have masses which are a tiny fraction of
this scale. This huge mass might be a hint that top quark plays an
essential role in the electroweak symmetry breaking. On the other
hand, the reported experimental data from Tevatron on the top
quark properties are still limited and no significant deviations
from the Standard Model predictions has been seen. Several
properties of the top quark have been already examined. They
consist of studies of the $t\bar{t}$ production cross section,
the top quark mass measurement, the measurement of $W$ helicity
in the top decay, the search for FCNC and many other studies
\cite{Tevatron}. However, it is expected that
top quark properties can be examined with high precision at the
LHC due to very large statistics \cite{Beneke}. Since the
dominant top quark decay mode is into a $W$ boson and a bottom
quark, the $tWb$ coupling can be investigated accurately. Within
the SM, the top quarks decay via electroweak interaction before
hadronization. This important property is one of the consequences
of its large mass. Hence, the spin information of the top quark
is transferred to its decay daughters and can be used as a
powerful mean for investigation of possible new physics.

There are many studies for testing the top quark decay properties
at hadron colliders. For instance, the non-standard effects on
the full top width have been investigated in the minimal
supersymmetric standard model and in the technicolor model
\cite{BSM0},\cite{BSM1},\cite{BSM2},\cite{BSM3},\cite{BH}. Some studies have been
performed on the effects of anomalous $tWb$ couplings on the top
width and some constraints have been applied on the anomalous
couplings
\cite{ATOP1},\cite{ATOP2},\cite{ATOP3},\cite{ATOP4},\cite{ATOP5},\cite{ATOP6}.

The noncommutativity in space-time is a possible generalization
of the usual quantum mechanics and quantum field theory to
describe the physics at very short distances of the order of the
Planck length, since the nature of the space-time changes at
these distances (motivations to construct models on
noncommutative space-time are coming from string theory, quantum
gravity, Lorentz breaking
\cite{Douglas1},\cite{Douglas2},\cite{Ardalan1},\cite{Ardalan2}).
In the simplest case, the noncommutativity in space-time is
described by a set of constant c-number parameters
$\theta^{\mu\nu}$ or equivalently by an energy scale
$\Lambda_{NC}$ and dimensionless parameters $C^{\mu\nu}$:

\begin{eqnarray}
[\hat{x}_{\mu},\hat{x}_{\nu}] = i\theta_{\mu \nu} = \frac{i}{\Lambda^{2}_{NC}}C_{\mu\nu}
= \frac{i}{\Lambda^{2}_{NC}}\left(
\begin{array}{cccc}
0 & -E_{1} & -E_{2} & -E_{3} \\
E_{1} & 0 & -B_{3} & B_{2}   \\
E_{2} & B_{3} & 0 & -B_{1}   \\
E_{3} & -B_{2} & B_{1} & 0  \\
\end{array}
\right)
\end{eqnarray}
where $\theta_{\mu \nu}$ is a real antisymmetric tensor with the
dimension of $[L]^{2}$. Here we have defined dimensionless
electric and magnetic parameters $(\vec{E},\vec{B})$ for
convenience. We note that a space-time noncommutativity,
$\theta_{0i}\neq 0$, might lead to some problems with unitarity
and causality \cite{Gomis},\cite{Chaichian}. It has been shown
that the unitarity can be satisfied for the case of
$\theta_{0i}\neq 0$ provided that $\theta^{\mu \nu}\theta_{\mu
\nu} > 0$ \cite{kostelecky}. However for simplicity, in this
article we take $\theta_{0i} = 0$ or equivalently $\vec{E} = 0$.

A noncommutative version of an ordinary field theory can be
obtained by replacing all ordinary products with Moyal $\star$
product defined as \cite{review}:
\begin{eqnarray}
(f\star g)(x) &=& \exp\left(\frac{i}{2}\theta^{\mu
\nu}\partial_{\mu}^{y}\partial_{\nu}^{z}\right)f(y)g(z)\bigg\vert_{y=z=x}\\
\nonumber &=&
f(x)g(x)+\frac{i}{2}\theta^{\mu\nu}(\partial_{\mu}f(x))(\partial_{\nu}g(x))+O(\theta^{2}).
\end{eqnarray}

The approach to the noncommutative field theory based on the Moyal
product and Seiberg-Witten maps allows the generalization of the
standard model to the case of noncommutative space-time, keeping
the original gauge group and particle content \cite{SW},\cite{Madore1},
\cite{Madore2},\cite{Madore3},\cite{Madore4},\cite{Reuter}.
Seiberg-Witten maps relate the noncommutative gauge fields and
ordinary fields in commutative theory via a power series
expansion in $\theta$. Indeed the noncommutative version of the
Standard Model is a Lorentz violating theory, but the Seiberg
Witten map shows that the zeroth order of the theory is the
Lorentz invariant Standard Model. The effects of noncommutative
space-time on some rare decay, collider processes, leptonic decay
of the $W$ and $Z$ bosons and additional phenomenological results
have been presented in \cite{OHL},
\cite{Haghighat},\cite{Schupp},\cite{Martin1},\cite{mojtaba},\cite{Josip1},
\cite{Josip2},\cite{Melic},\cite{Namit},\cite{Iltan},\cite{arfaei}
and some limits have been set on noncommutative scale.

The aim of this article is to estimate the bounds on the
noncommutative scale $\Lambda_{NC}$ and to estimate the $95\%$
exclusion contours for $\vec{B}$ using the current measurements of
the top quark width and the $W$ boson polarization in the
$t\bar{t}$ events from CDF experiment at Tevatron. In Section 2, a
short introduction for the noncommutative standard model (NCSM) is given.
Section 3 is dedicated to review the $W$ boson polarization in the top quark events.
Section 4 presents the noncommutative effects on the top quark
width and limit on $\Lambda_{NC}$ from current measured top
width. Section 5 gives the limits on $\Lambda_{NC}$ and $\vec{B}$
using $W$ boson polarization . Finally, Section 6 concludes the
paper.

\section{The noncommutative standard model (NCSM)}
The action of the NCSM can be obtained by replacing the ordinary
products in the action of the classical SM by the Moyal products
and then matter and gauge fields are replaced by the appropriate
Seiberg-Witten expansions. The action of NCSM can be written as:
\begin{eqnarray}
S_{NCSM} = S_{fermions} + S_{gauge} + S_{Higgs} + S_{Yukawa},
\end{eqnarray}
This action has the same structure group $SU(3)_{C}\times
SU(2)_{L}\times U(1)_{Y}$ and the same fields number of coupling
parameters as the ordinary SM. The approach which has been used in
\cite{Madore1},\cite{Madore2},\cite{Madore3},\cite{Madore4} to
build the NCSM is the only known approach that allows to build
models of electroweak sector directly based on the structure group
$SU(2)_{L}\times U(1)_{Y}$ in a noncommutative background. The
NCSM is an effective, anomaly free, noncommutative field theory
\cite{Martin2},\cite{Martin3}.

We just consider the fermions (quarks and leptons). The fermionic
matter part in a very compact way is:
\begin{eqnarray}
S_{fermions} = \int d^{4}x
\sum_{i=1}^{3}\left(\bar{\widehat{\Psi}}^{(i)}_L\star
(i\widehat{\DS} ~\widehat{\Psi}^{(i)}_L)\right) + \int d^{4}x
\sum_{i=1}^{3}\left(\bar{\widehat{\Psi}}^{(i)}_R\star
(i\widehat{\DS} ~\widehat{\Psi}^{(i)}_R)\right),
\end{eqnarray}
where $i$ is generation index and $\Psi^{i}_{L,R}$ are:
\begin{eqnarray}
\Psi^{(i)}_L = \left(
                 \begin{array}{c}
                   L^{i}_{L} \\
                   Q^{i}_{L} \\
                 \end{array}
               \right)
~,~\Psi^{(i)}_R = \left(
                 \begin{array}{c}
                   e^{i}_{R} \\
                   u^{i}_{R} \\
                   d^{i}_{R}
                 \end{array}
               \right)
\end{eqnarray}
where $L^{i}_{L}$ and $Q^{i}_{L}$ are the well-known lepton and
quark doublets, respectively. The Seiberg-Witten maps for the
noncommutative fermion and vector fields yield:
\begin{eqnarray}
\widehat{\psi} &=& \widehat{\psi}[V] = \psi -
\frac{1}{2}\theta^{\mu\nu}V_{\mu}\partial_{\nu}\psi +
\frac{i}{8}\theta^{\mu\nu}[V_{\mu},V_{\nu}]\psi + O(\theta^{2}),\nonumber \\
\widehat{V_{\alpha}} &=& \widehat{V_{\alpha}}[V] = V_{\alpha} +
\frac{1}{4}\theta^{\mu\nu} \{\partial_{\mu}V_{\alpha} +
F_{\mu\alpha}, V_{\nu}\} + O(\theta^{2}),
\end{eqnarray}
where $\psi$ and $V_{\mu}$ are ordinary fermion and gauge fields,
respectively. Noncommutative fields are denoted by a hat. For a
full description and review of the NCSM, see
\cite{Madore1},\cite{Madore2},\cite{Madore3},\cite{Madore4}. The
$t(p_{1})\rightarrow W(q)+b(p_{2})$ vertex in the NCSM up to the
order of $\theta^{2}$ can be written as
\cite{mojtaba},\cite{Namit}:
\begin{eqnarray}\label{vertex}
\Gamma_{\mu,NC} &=& \frac{g V_{tb}}{\sqrt{2}}[\gamma_{\mu} +
\frac{1}{2}(\theta_{\mu\nu}\gamma_{\alpha}+\theta_{\alpha\mu}\gamma_{\nu}+
\theta_{\nu\alpha}\gamma_{\mu})q^{\nu}p^{\alpha}_{1} \\
\nonumber
&-&\frac{i}{8}(\theta_{\mu\nu}\gamma_{\alpha}+\theta_{\alpha\mu}\gamma_{\nu}+
\theta_{\nu\alpha}\gamma_{\mu})(q\theta
p_{1})q^{\alpha}p^{\nu}_{1}]P_{L}.
\end{eqnarray}
where $P_{L} = \frac{1-\gamma_{5}}{2}$ and $q\theta p_{1}\equiv
q^{\mu}\theta_{\mu\nu}p^{\nu}_{1}$. This vertex is similar to the
vertex of $W$ decays into a lepton and anti-neutrino \cite{Iltan}.
However, one should note that due to the ambiguities in the SW maps
there are additional terms in the above vertex.  
Since they will not affect the results, we have ignored them\cite{Ruckl}.

\section{$W$ boson polarization in top events}
This section presents the observables used to measure the
polarization of the $W$ boson. The real $W$ in the $t\rightarrow
W+b$ decay can be produced with a longitudinal, left-handed or
right-handed helicity. The corresponding probabilities are $F_{0},
F_{L}$ and $F_{R}$, respectively, whose SM expectations at tree
level in the zero b-mass approximation are:

\begin{eqnarray}
F_{0} &=& \frac{\Gamma(t\rightarrow W_{0}b)}{\Gamma(t\rightarrow
Wb)} = \frac{m^{2}_{t}}{m^{2}_{t}+2m^{2}_{W}} = 0.703 \\
\nonumber F_{L} &=& \frac{\Gamma(t\rightarrow
W_{L}b)}{\Gamma(t\rightarrow Wb)} =
\frac{2m^{2}_{W}}{m^{2}_{t}+2m^{2}_{W}} = 0.297\\ \nonumber
 F_{R} &=&
\frac{\Gamma(t\rightarrow W_{R}b)}{\Gamma(t\rightarrow Wb)} =
0.000
\end{eqnarray}

where $m_{t}$ and $m_{W}$ are the top and $W$ masses in GeV.
$\Gamma(t\rightarrow Wb)$ is the top quark width. We have the
restriction $F_{0} + F_{L} + F_{R}$ = 1. Since massless particles
must be left-handed in the SM, right-handed $W$ bosons do not
exist in the zero b-mass approximation, due to angular momentum
conservation. Including QCD and electroweak radiative corrections,
finite width corrections and non-zero b-quark mass induces small
variations: $F_{0} = 0.695, F_{L} = 0.304$ and $F_{R}$ = 0.001 for
$m_{t}$ = 175 GeV/c$^{2}$ \cite{wcorr}. Because the top quark is
very heavy, $F_{0}$ is large and the top decay is the only
significant source of longitudinal W bosons. Deviations of
$F_{0}$ from its SM value would bring into question the validity
of the Higgs mechanism of the spontaneous symmetry breaking,
responsible for the longitudinal degree of freedom of the massive
gauge bosons. Any deviation of $F_{R}$ from zero could point to a
non-Standard Model couplings such as $Wtb$ anomalous couplings or
new couplings coming from space-time noncommutativity introduced
in the last section. The best way to access particle spin
information is to measure the angular distribution of its decay
products, thereby called spin analyzers. As an example, the
charged lepton from the decay of longitudinally polarized $W$
boson tends to be emitted transversally to the $W$ boson
direction, due to angular momentum conservation. Similarly, the
charged lepton from a left-handed (right-handed) $W$ boson is
preferentially emitted in the opposite (same) $W$ boson
direction. By definition of $\theta^{\ast}_{l}$ to be the angle
between the charged lepton direction in the $W$ boson rest frame
and the $W$ direction in the top quark rest frame, the normalized
differential decay rate can be expressed as the following
\cite{Kane}:
\begin{eqnarray}
\frac{1}{\Gamma}\frac{d\Gamma}{d\cos\theta^{\ast}_{l}} =
\frac{3}{8}(1+\cos\theta^{\ast}_{l})^{2}F_{R}+\frac{3}{8}(1-\cos\theta^{\ast}_{l})^{2}F_{L}+
\frac{3}{4}(\sin\theta^{\ast}_{l})^{2}F_{0}.
\end{eqnarray}

The measured values of the fractions $F_{0}$ and $F_{R}$ of
longitudinally polarized and right-handed $W$ bosons in top quark
decays using data collected with the CDF II detector(the data set
used in the analysis corresponds to an integrated luminosity of
approximately 955 pb$^{-1}$) are as the following \cite{WPOLCDF}:
\begin{eqnarray}
F_{0} &=& 0.59 \pm 0.12(\text{stat})^{+0.07}_{-0.06}(\text{syst})\\ \nonumber
F_{R} &=& -0.03 \pm 0.06(\text{stat})^{+0.04}_{-0.03}(\text{syst})\\ \nonumber
F_{R} &\leq& 0.1~~ \text{at 95$\%$ Confidence Level}
\end{eqnarray}

\section{The noncommutative effects on the top quark width}
Using the introduced Feynman rule for the $Wtb$ vertex in the
NCSM in eq.\ref{vertex}, the decay rate is easily evaluated. The
decay rate in the top quark rest frame, which contains the noncommutative effects can be
expressed as the following \cite{Namit}:
\begin{eqnarray}\label{wtop}
\Gamma(t\rightarrow Wb) = \frac{|V_{tb}|^{2}}{16\pi
m_{t}}(\frac{g}{2\sqrt{2}m_{W}})^{2}\lambda^{1/2}(1,\frac{m^{2}_{W}}{m^{2}_{t}},\frac{m^{2}_{b}}{m^{2}_{t}})
[A_{SM}+A_{NC}]
\end{eqnarray}
where,
\begin{eqnarray}
A_{SM} &=&
2[m^{2}_{W}(m^{2}_{t}+m^{2}_{b}-m^{2}_{W})+((m^{2}_{t}-m^{2}_{b})^{2}-m^{4}_{W})]\\
\nonumber A_{NC} &=&
\frac{m^{2}_{W}}{12\Lambda^{4}_{NC}}|\vec{B}|^{2}(5m^{2}_{b}+m^{2}_{t}-m^{2}_{W})[m^{4}_{b}+(m^{2}_{t}-m^{2}_{W})^{2}-2m^{2}_{b}(m^{2}_{t}+m^{2}_{W})]\nonumber
\end{eqnarray}

One should note that in the above relation we have set
$\theta_{i0} = 0$ and the terms with $\theta_{ij}$ or
equivalently $\vec{B}$ are kept.

The SM prediction for the top quark lifetime is around $4\times
10^{-25}$ s which corresponds to the top quark width of 1.5 GeV.
It is notable that because of the limited resolutions of the
experiments, it is very difficult to measure this very short
lifetime or the corresponding width. However, we are able to set
an upper limit on the top quark width from the available data from
Tevatron. In \cite{TopWidthCDF} an upper limit has been set on the
top quark width using a likelihood fit to the reconstructed top
mass distribution. In the analysis the lepton+jets channel of
$t\bar{t}$ candidates, in which one of two W-bosons decays to
$l\nu_{l}$ while the other decays to $q\bar{q}$, is used to
reconstruct the top quark mass. Finally, the estimated upper bound
on the top quark width is 12.7 GeV with $95\%$ C.L. This is
corresponding to the lower limit of $5.2\times 10^{-26}$ s for
the top quark lifetime.

The measured upper limit on the top quark width and eq.\ref{wtop}
lead to the following bound on $\Lambda_{NC}$:
\begin{eqnarray}
\Lambda_{NC} \geq 624~\text{GeV for }|\vec{B}|^{2} = 1
~~\text{with $95\%$ C.L.}
\end{eqnarray}

\section{The effects of noncommutativity on $W$ boson polarization}
The noncommutative corrections to $F_{0}$ and $F_{R}$ can be
calculated using the general $Wtb$ vertex given by
eq.\ref{vertex}. The noncommutative corrections to $F_{0}$ are
proportional to $|\theta_{i0}|^{2}$ and there is no contribution
from $|\theta_{ij}|$. Therefore $F_{0}$ can not provide us any
information about $|\theta_{ij}|$.

The noncommutative corrections to $F_{R}$ can be evaluated by
calculation the corrections to $\Gamma(t\rightarrow W_{R}b)$:
\begin{eqnarray}
\Gamma_{NC}(t\rightarrow W_{R}b) &=&\frac{|V_{tb}|^{2}}{16\pi
m_{t}}(\frac{g}{2\sqrt{2}})^{2}\lambda^{1/2}(1,\frac{m^{2}_{W}}{m^{2}_{t}},\frac{m^{2}_{b}}{m^{2}_{t}})\Delta_{R}\\
\Delta_{R} &=&
\frac{m^{4}_{t}}{48}(m^{2}_{t}+m^{2}_{b}-m^{2}_{W})\times\sum^{2}_{i=1}\theta^{i\mu}\theta^{i}_{\mu}
\nonumber
\end{eqnarray}

and $F_{R}$ is:
\begin{eqnarray}
F_{R} = \frac{\Gamma_{NC}(t\rightarrow
W_{R}b)}{\Gamma(t\rightarrow Wb)}
\end{eqnarray}
In obtaining the above relation we have neglected the
contributions from $\theta_{i0}$. The sum over $i$ is from one to two which 
is due the fact that the top quark is only restricted to decay into right-handed W-boson.
Obviously, even in the limit of
vanishing b-quark mass and neglecting the QCD and electroweak
corrections, there are non-zero contributions to the $F_{R}$ from
noncommutativity. $F_{R}$ is
very sensitive to $\Lambda_{NC}$ for low values. The maximum value of $F_{R}$ which is equal to one 
corresponds to $\Lambda_{NC} \sim 870$ GeV. With the
increase in $\Lambda_{NC}$, $F_{R}$ approaches the SM value which is zero. The
combination of the upper limit on $F_{R}$ measured by CDF,
mentioned in Section 3, and the above relation leads to:
\begin{eqnarray}
\Lambda_{NC} \geq 1550~\text{GeV for }|\vec{B}|^{2} = 1
~~\text{with $95\%$ C.L.}
\end{eqnarray}
This bound is higher than the one obtained in \cite{OHL} (from
$Z\gamma$ production at the Tevatron and the LHC) and the bound
obtained in \cite{Josip2} from SM forbidden decays which is
$\Lambda_{NC} > 1$ TeV. It is noticeable that any better
measurement on the upper bound of $F_{R}$ causes to a higher
limit on $\Lambda_{NC}$.

\begin{figure}
\centering
  \includegraphics[width=7cm,height=6cm]{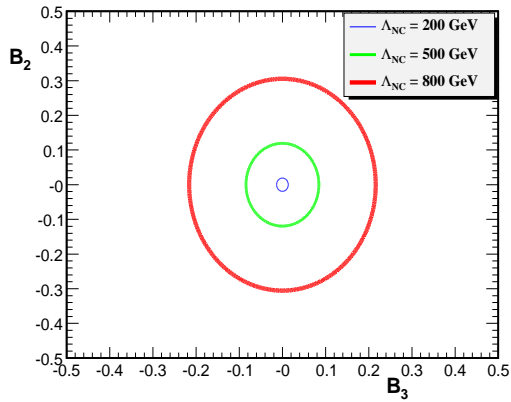}\\
  \caption{The $95\%$ C.L. exclusion contours
on $B_{2},B_{3}$ for different values of
$\Lambda_{NC}$.}\label{contours}
\end{figure}

The noncommutative corrections to $F_{R}$ and the CDF upper also
limit provides the $95\%$ C.L. exclusion contours on
$B_{1},B_{2},B_{3}$ for different values of $\Lambda_{NC}$
(200,500,800 GeV). These contours are presented in Figure
\ref{contours}.
According to the form of variation of $\vec{B}$ with changing the frame (such as the
Lorentz transformation of magnetic field in electrodynamics),
the change of frame leads to slightly lower limits. 

The estimated bounds are comparable with the bounds
estimated in \cite{OHL} and lead to: $|\theta_{ij}|\leq 10^{-7}
\text{GeV}^{-2}$.

%\begin{table}
%\begin{center}
%\begin{tabular}{|c|c|c|c|}\hline
%   NC Parameter          &  $B_{1}$      &  $B_{2}$   &  $B_{3}$    \\\hline\hline
%    $\Lambda_{NC} = 200$ GeV &$-0.02\leq B_{1} \leq +0.02 $ &  $-0.02\leq B_{2} \leq +0.02 $ & $-0.014\leq B_{3} \leq +0.014 $  \\\hline
%   $\Lambda_{NC} = 500$ GeV &$-0.12\leq B_{1} \leq +0.12 $  &  $-0.12\leq B_{2} \leq +0.12 $ & $-0.08\leq B_{3} \leq +0.08 $  \\\hline
%   $\Lambda_{NC} = 800$ GeV &$-0.31\leq B_{1} \leq +0.31 $  &  $-0.31\leq B_{2} \leq +0.31 $ & $-0.22\leq B_{3} \leq +0.22 $  \\
%   \hline
%   \end{tabular}\label{bounds}
%\end{center}\caption{The $95\%$ C.L. limits on the $B_{1},B_{2},B_{3}$
%corresponding to $\Lambda_{NC} = 200~,500~,800$ GeV.}
%\end{table}

\section{Conclusion}

The recent measurements of the top quark width and $W$ boson
polarization have been used to estimate the noncommutative scale
$\Lambda_{NC}$ and to estimate the $95\%$ exclusion contours on
$B_{1,2,3}$. The extracted limits confirms the limits obtained by
other studies . The $95\%$ bound on $\Lambda_{NC}$ from the
measured top quark width and $W$ polarization are $\Lambda_{NC}
\geq 625$ GeV, $\Lambda_{NC} \geq 1550$ GeV (assuming
$|\vec{B}|^{2} = 1$), respectively. The obtained limit on
$\Lambda_{NC}$ from $W$ boson polarization is higher than the one
obtained from the top width and also is higher than the limits
obtained in \cite{OHL} from $Z\gamma$ production and the limits
estimated in \cite{Josip2} from SM forbidden decays. The
$95\%$ exclusion contours on $B_{1,2,3}$ lead to $|\theta_{ij}|\leq 10^{-7} \text{GeV}^{-2}$.
\\

{\large \bf Acknowledgments}\\
The author would like to thank Prof. H. Arfaei for fruitful discussions.
\\

\end{document}